\documentclass{article}
\usepackage{amsmath,amssymb}
\usepackage{graphicx}
\usepackage{oldlfont}
\usepackage{stmaryrd}
\usepackage{colordvi}
\usepackage{epsfig}

\usepackage{amsfonts}
\usepackage{amssymb}
\usepackage{amscd}

\usepackage{stmaryrd}

\newcommand{\ar}{{\mathfrak r}} 
\newcommand{\deq}{\doteq}

\newcommand{\pfont}{\sffamily\bfseries}
\newcommand{\save}[1]{}

\renewcommand{\leq}{\leqslant}
\renewcommand{\geq}{\geqslant}

\newcommand{\li}{\item\zero}

\newcommand{\xfig}[2]{\medskip \centerline{\epsfig{figure=#1.eps,height=#2}} \bigskip}

\newcommand{\ul}[1]{\underline{#1}}

\newcommand{\ara}[1]{{\stackrel{#1}{\longrightarrow}}} 

\newcommand{\dara}[2]%
      {\overset{#1}{\underset{\raisebox{4ex}{\;\;\fnit{#2}}}{\longrightarrow}}} 

\newcommand{\mywhite}[1]{\White{#1}}

\newcommand{\outt}[1]{}
\newcommand{\delete}[1]{}
\newcommand{\blank}[1]{}

\newcommand{\notedomission}[1]{\medskip\noindent{\bf TEXT OMITTED}\\[2mm]}

\newenvironment{xtr}{\huge\tt\hspace{-5mm}}{\\[2mm] \zero\hrulefill} 
\newcommand{\bxtr}{\begin{xtr}\begin{envviolet}\noindent}
\newcommand{\extr}{\end{envviolet}\end{xtr}}

\newcommand{\bmini}{\begin{minipage}[t]{0.8\textwidth}}
\newcommand{\emini}{\end{minipage}}

\newenvironment{envviolet}{\textViolet}{\textBlack}

\newcommand{\bxit}[1]{\hbox{\it #1}}

\newcommand{\bxrm}[1]{\hbox{\rm #1}}

\newcommand{\bxbf}[1]{\hbox{\bf #1}}

\newcommand{\bxsc}[1]{\hbox{\sc #1}}
\newcommand{\bxtt}[1]{\hbox{$\tt #1$}}

\newcommand{\bxsl}[1]{\hbox{\sl #1}}

\newcommand{\fnit}[1]{\hbox{\footnotesize\it #1}}

\newcommand{\ttscript}[1]{\hbox{\scriptsize\bf #1}}

\newcommand{\bfc}{\hbox{\bf c}}

\newcommand{\bff}{\hbox{\bf f}}

\newcommand{\bfq}{\hbox{\bf q}}

\newcommand{\bft}{\hbox{\bf t}}

\newcommand{\sla}{\hbox{$\sl a$}}

\newcommand{\tta}{\hbox{$\tt a$}}
\newcommand{\ttb}{\hbox{$\tt b$}}
\newcommand{\ttc}{\hbox{$\tt c$}}
\newcommand{\ttd}{\hbox{$\tt d$}}
\newcommand{\tte}{\hbox{$\tt e$}}
\newcommand{\ttf}{\hbox{$\tt f$}}
\newcommand{\ttg}{\hbox{$\tt g$}}

\newcommand{\ttp}{\hbox{$\tt p$}}
\newcommand{\ttq}{\hbox{$\tt q$}}
\newcommand{\ttr}{\hbox{$\tt r$}}
\newcommand{\tts}{\hbox{$\tt s$}}
\newcommand{\ttt}{\hbox{$\tt t$}}
\newcommand{\ttu}{\hbox{$\tt u$}}

\newcommand{\ttz}{\hbox{$\tt z$}}

\newcommand{\ttzero}{\hbox{$\tt 0$}}

\newcommand{\ttone}{\hbox{$\tt 1$}}

\newcommand{\bfI}{\hbox{\bf I}}

\newcommand{\bfN}{\hbox{\bf N}}

\newcommand{\gothA}{\mbox{$\mathfrak A$}}

\newcommand{\gothC}{\mbox{$\mathfrak C$}}

\newcommand{\gothF}{\mbox{$\mathfrak F$}}

\newcommand{\calN}{\hbox{$\cal N$}}

\newcommand{\calQ}{\hbox{$\cal Q$}}

\newcommand{\calS}{\hbox{$\cal S$}}
\newcommand{\calT}{\hbox{$\cal T$}}

\newcommand{\tinycalQ}{\hbox{{\tiny $\cal Q$}}}

\newcommand{\tinycalS}{\hbox{{\tiny $\cal S$}}}
\newcommand{\tinycalT}{\hbox{{\tiny $\cal T$}}}

\newcommand{\ttE}{\hbox{$\tt E$}}

\newcommand{\dN}{\hbox{$\mathbb N$}}

\newcommand{\gra}{\hbox{$\alpha$}}
\newcommand{\grb}{\hbox{$\beta$}}
\newcommand{\grg}{\hbox{$\gamma$}}
\newcommand{\grd}{\hbox{$\delta$}}
\newcommand{\gre}{\hbox{$\varepsilon$}}
\newcommand{\grh}{\hbox{$\eta$}}
\newcommand{\grz}{\hbox{$\zeta$}}

\newcommand{\grl}{\hbox{$\lambda$}}

\newcommand{\grn}{\hbox{$\nu$}}
\newcommand{\grx}{\hbox{$\xi$}}

\newcommand{\grs}{\hbox{$\sigma$}}
\newcommand{\grt}{\hbox{$\tau$}}

\newcommand{\grf}{\hbox{$\varphi$}}

\newcommand{\grq}{\hbox{$\psi$}}

\newcommand{\grG}{\hbox{$\Gamma$}}
\newcommand{\grD}{\hbox{$\Delta$}}

\newcommand{\grP}{\hbox{$\Pi$}}
\newcommand{\grS}{\hbox{$\Sigma$}}

\newcommand{\grF}{\hbox{$\Phi$}}

\newcommand{\bemptyset}{\hbox{$\pmb \emptyset$}}

\newcommand{\bfgrw}{\hbox{$\pmb \omega$}}

\newcommand{\bgrw}{\hbox{\boldmath$\omega$}}

\newcommand{\sbwedge}{\hbox{$\bigwedge$}} 
\newcommand{\sbvee}{\hbox{$\bigvee$}}
\newcommand{\sbcup}{\hbox{$\bigcup$}}

\newcommand{\ra}{\rightarrow}
\newcommand{\pa}{\rightharpoonup}
\newcommand{\sra}{\!\rightarrow\!} 

\newcommand{\la}{\leftarrow}
\newcommand{\rA}{\Rightarrow}  

\newcommand{\lra}{\leftrightarrow}

\newcommand{\splus}{\!+\!}
\newcommand{\sminus}{\!-\!}
\newcommand{\seq}{\!=\!}

\newcommand{\suparrow}{\neghalfmm\uparrow\neghalfmm}

\newcommand{\ssUparrow}{\negonemm\Uparrow\negonemm}
\newcommand{\sdownarrow}{\neghalfmm\downarrow\neghalfmm}
\newcommand{\sDownarrow}{\neghalfmm\Downarrow\neghalfmm}
\newcommand{\ssDownarrow}{\negonemm\Downarrow\negonemm}

\newcommand{\qed}{\hfill {\boldmath$\Box$}\\}
\newcommand{\lng}{\langle}
\newcommand{\rng}{\rangle}
\newcommand{\df}{=_{\rm df}}

\newcommand{\rsem}{\,]\hspace{-0.6mm}]} 
\newcommand{\lsem}{[\hspace{-0.6mm}[\,} 

\newcommand{\ignore}[1]{}

\newcommand{\mx}{\makebox}

\newcommand{\zero}{\rule{0mm}{3mm}}

\newcommand{\negonemm}{\mx[-1mm]{}}
\newcommand{\neghalfmm}{\mx[-0.5mm]{}}

\newcommand{\minusthreecm}{\mx[-3cm]{}}

\newcommand{\onecm}{\mx[1cm]{}}
\newcommand{\twocm}{\mx[2cm]{}}

\newcommand{\fivecm}{\mx[5cm]{}}

\newcommand{\bc}{\begin{center}}
\newcommand{\ec}{\end{center}}
\newcommand{\beq}{\begin{equation}}
\newcommand{\eeq}{\end{equation}}

\newcommand{\be}{\begin{enumerate}}
\newcommand{\ee}{\end{enumerate}}
\newcommand{\bi}{\begin{itemize}}
\newcommand{\ei}{\end{itemize}}
\newcommand{\bd}{\begin{description}}
\newcommand{\ed}{\end{description}}
\newcommand{\beqn}{\begin{equation}}
\newcommand{\eeqn}{\end{equation}}

\newcommand{\beqna}{\begin{eqnarray}}
\newcommand{\eeqna}{\end{eqnarray}}
\newcommand{\beqnas}{\begin{eqnarray*}}
\newcommand{\eeqnas}{\end{eqnarray*}}
\newcommand{\beqnaa}{$$\begin{array}{rcll}}  
\newcommand{\eeqnaa}{\end{array}$$}  
\newcommand{\beqnal}{$$\begin{array}{l}}  
\newcommand{\eeqnal}{\end{array}$$}  
\newcommand{\beqnana}{$$\begin{array}{lrcll}}  
\newcommand{\eeqnana}{\end{array}$$}  
\newcommand{\btbl}[1]{\begin{center}\begin{tabular}{#1}}
\newcommand{\etbl}{\end{tabular}\end{center}}
\newcommand{\beqnc}{$$\begin{array}{rclcl}}
\newcommand{\eeqnc}{\end{array}$$}
\newcommand{\fn}{\footnote}

\newcommand{\Section}[1]{{\section{{\pfont #1}}}}
\newcommand{\prf}{{\sc Proof. }}

\newtheorem{dclprop}{{\sc Proposition}} 
\newtheorem{dclprops}{{\sc Proposition}}[subsection] 
\newtheorem{dclbigthm}[dclprop]{THEOREM}

\makeatletter
\def\thmlabel#1{\@bsphack\if@filesw {\let\thepage\relax
\xdef\@gtempa{\write\@auxout{\string
\newlabel{#1}{{\@Roman{\@currentlabel}}{\thepage}}}}}\@gtempa
\if@nobreak \ifvmode\nobreak\fi\fi\fi\@esphack}
\makeatother

\newtheorem{dclthm}[dclprop]{{\sc Theorem}}   
\newtheorem{dclthms}[dclprops]{{\sc Theorem}}   

\newtheorem{dcllem}[dclprop]{{\sc Lemma}}
\newtheorem{dcllems}[dclprops]{{\sc Lemma}} 
\newtheorem{dclsublem}[dclprop]{{\sc Sublemma}}
\newtheorem{dclcor}[dclprop]{{\sc Corollary}}
\newtheorem{dclcors}[dclprops]{{\sc Corollary}} 
\newtheorem{dcldfn}[dclprop]{{\sc Definition}}
\newtheorem{dcldfns}[dclprops]{{\sc Definition}}
\newtheorem{dclasss}[dclprops]{{\bf Assumption}}
\newtheorem{dclass}[dclprop]{{\bf Assumption}}

\newenvironment{prop}{\medskip\begin{dclprop}\sl}{\end{dclprop}}
\newenvironment{thm}{\medskip\begin{dclthm}\sl}{\end{dclthm}}

\newenvironment{cor}{\medskip\begin{dclcor}\sl}{\end{dclcor}}

\newenvironment{dfn}{\medskip\begin{dcldfn}\sl}{\end{dcldfn}}

\newenvironment{lem}{\medskip\begin{dcllem}\sl}{\end{dcllem}}

\newcommand{\bsl}{\begin{verse}\sl}
\newcommand{\esl}{\end{verse}}

\newtheorem{ex}[dclprop]{Example}
\newtheorem{exxs}[dclprop]{Exercises}

\newenvironment{exercises-with-preamble}{\begin{exxs}\rm}{\end{exxs}}

\newcommand{\bthm}{\begin{thm}}
\newcommand{\ethm}{\end{thm}}
\newcommand{\bprop}{\begin{prop}}
\newcommand{\eprop}{\end{prop}}
\newcommand{\blem}{\begin{lem}}
\newcommand{\elem}{\end{lem}}
\newcommand{\bcor}{\begin{cor}}
\newcommand{\ecor}{\end{cor}}
\newcommand{\bdfn}{\begin{dfn}}
\newcommand{\edfn}{\end{dfn}}

\newcommand{\bz}{\begin{quote}\small}
\newcommand{\ez}{\end{quote}}

\newcommand{\einference}[2]  
  {\shortstack
      {$ #1 $\\ \mbox{}\\ $ #2 $}}

\newlength{\txtlth}
\newlength{\txtht}

\newcommand{\vsup}[2]{\vec{#1}\raisebox{0.7ex}{{\tiny\it #2}}} 

\renewcommand{\sla}{\! \leftarrow \!}
\newcommand{\bexample}{\begin{ex}}
\newcommand{\eexample}{\end{ex}}

\newcommand{\btu}{{\vartriangle}}
\newcommand{\btd}{%
    \mathrel{\reflectbox{\rotatebox[origin=c]{180}{{{\scriptsize $\vartriangle$}}}}}}

\begin{document}
\title{{\pfont A theory of finite structures}}
\author{\bxit{Daniel Leivant}\\
{\normalsize{SICE (Indiana University) and IRIF (Paris-Diderot)}}}

\maketitle

\begin{abstract}
We develop a novel formal theory of finite structures,
based on a view of finite structures as
a fundamental artifact of computing and programming, forming
a common platform for computing both within particular finite structures, 
and in the aggregate for computing over infinite data-types
construed as families of finite structures.
A {\em finite structure} is here a finite collection of 
finite partial-functions,
over a common universe of {\em atoms}.  
The theory is second-order, as it
uses quantification over finite functions.

Our formal theory \bxbf{FS} uses a small number of fundamental
axiom-schemas, with finiteness enforced by a schema of
induction on finite partial-functions.  We show that computability
is definable in the theory by existential formulas, generalizing
Kleene's Theorem on the $\grS_1$-definability of RE sets,
and use that result to prove that \bxbf{FS} is mutually interpretable
with Peano Arithmetic.
\end{abstract}

\Section{Introduction}

We develop a formal theory of finite structures, motivated in part
by an imperative programming language for transforming such structures.  

Our point of departure is to posit finite structures as 
a fundamental artifact of computing and programming.
It is a truism that a database is a finite structure. 
But elements of infinite data-types, such as the natural numbers
or binary strings, are also finite-structures that obey certain
requirements.
Computing over such a data-type can thus be construed as
a transformation process driven by those structures'
internal making.
Viewed from that angle, 
a function computed by a transducer-program is perceived 
as a mapping over the space of finite-structures, rather than
a function {\em within} a particular infinite structure.
Thus, finite partial-structures form a common platform
for computing both within particular finite structures, and in the aggregate
for computing over infinite data-types, as long as their elements
are not infinite themselves (e.g.\ streams).

Our theory is second-order, in that we 
quantify over structures (via a quantification over finite functions).
The well-known second-order nature of inductive data is thus manifested
in the computation objects being {\em themselves} second-order, 
albeit finite ones.
The atoms, from which our finite structures are built, 
are the first-order elements. They are nameless and structure-less, and
there must be an unlimited supply of them to permit unhindered
structure extension during computation.
One intends to identify structures
that differ only by the choice of atoms, i.e.\ ones that are isomorphic
to each other,
though it turns out that we can often convey that intent implicitly,
without complicating matters with permanent references to equivalence classes.

Since referencing objects by constant and function identifiers
is central to imperative programming, we choose to base
our structures on partial-functions, rather
than sets or relations.
That is, each one of our structures is a finite set of finite 
partial-functions. 
The Tarskian notion of an explicit structure universe is superfluous here:
the only atoms that matter are the ones
that appear as inputs and/or outputs of a structure's functions, 
a set we shall call the structure's {\em scope.}

Finally, as we base our finite structures on finite partial-functions,
the basic operations on structures must be function-updates of
some form.  This is analogous to the operation of adjoining an element 
to a set, which underlies several existing theories of finite sets (see below).

In fact, the idea that inductive data-objects, such as the natural numbers,
can be construed as being composed of underlying units, 
goes back all the way to
Euclid, who defined a number as ``a multitude composed of units"
\cite[7th book, definition 2]{EuclidElements}.
The logicist project 
of Frege, Dedekind and Russell, which attempted to reduce mathematics
to logic's first principles, included this very same reduction of
natural numbers to finite sets.
The current predominant view of natural numbers
as irreducible primal objects
was advocated by opponents of the logicist project,
notably Poincar\'{e} \cite{Goldfarb88} and Kronecker,
whose critique was subsequently advanced by the emergence
of Tarskian semantics and by G\"{o}del's incompleteness theorems.


Philosophical considerations aside, the kinship between natural numbers
and finite sets raises interesting questions about the formalization
of finite set theory, and the mutual
interpretation of such a theory with formal theories for arithmetic.
Several formal theories proposed for finite set theory are based
on \bxbf{ZF} with the Axiom of Infinity replaced by its negation,
which we denote here by $\bxbf{ZF}^f$.
\cite{KayeW07} shows that $\bxbf{ZF}^f$ can be interpreted in
\bxbf{PA} (Theorem 3.1), and that \bxbf{PA} can be interpreted in 
$\bxbf{ZF}^f$ (Theorem 4.5),
but not by the inverse of the former.  That inverse would work
for the interpretation of \bxbf{PA} provided $\bxbf{ZF}^f$ is 
augmented with a {\em Transitive Containment} axiom, asserting
that every set is contained in a transitive set (Theorem 6.5).
A result analogous to the latter\fn{Wang removes from \bxbf{ZF}
Infinity without replacing it with its negation,
since he does not interpret finite set theory in {\bf PA}.}
was proved by Wang already in 1953 \cite{Wang53}, 
where he uses an axiom (ALG) analogous to Transitive Containment,
stating the enumerability of the collection of all finite sets.

It seems, though, that $\bxbf{ZF}^f$ is somewhat of an oxymoron, in that 
it embraces \bxbf{ZF}, a theory designed specifically 
to reason about infinity, only to eviscerate it off the bat
by excluding infinity.\fn{The subtext is, of course, the conviction that
\bxbf{ZFC} is the canonical formal framework for all of mathematics.}
An approach more germane to finiteness was proposed already
by Zermelo \cite[Theorem 3]{Zermelo09}, 
Whitehead and Russell  \cite[*120.23]{WhiteheadRussell12},
Sierpi\'{n}ski \cite[p.106]{Sierpinski19},
Kuratowski \cite{Kuratowski20}, and Tarski
\cite{Tarski24}.
It enforces finiteness by an induction
principle for sets: if $\emptyset$ satisfies a property \grf\
and whenever $X$ satisfies \grf\ then so does
$X \cup \{y\}$ for each $y$, then all sets satisfy \grf. 
This induction principle corresponds to the inductive definition of finite sets
as the objects generated by successively 
joining elements to $\emptyset$.

Mayberry \cite{Mayberry11} defines an arithmetic based on induction on sets,
which he deftly dubs {\em Euclidean Arithmetic.}
He considers a number of induction principles derived from the
set-induction principle above, but 
allows for neither set quantifiers nor unbounded atomic quantifiers
in the induction formula.
Consequently, his system is mutually interpretable
with a weak number theory,
namely $I\Delta_0$ extended with an exponentiation function
(detailed proof in \cite{Homolka83}).
Pettigrew \cite{Pettigrew08,Pettigrew09}
shows that these two theories are also equipollent with 
a bounded version of $\bxbf{ZF}^f$.


Our approach differs from the ones above in several respects.
Our theory \bxbf{FS} of finite structures
is based on finite structures rather than finite sets, 
and is formally second-order, in contrast to theories based on
\bxbf{ZF}, which on the one hand are first-order, and
on the other hand refer to a cumulative hierarchy of sets.
At the same time, \bxbf{FS}'s second-order nature is weak, in that
its induction principle implies that its second-order object
are finite.
As a result, our theory of finite structures, although based
on a variant of set induction (and not $\bxbf{ZF}^f$) is equipollent to
full Peano Arithmetic.

Principally, our motivation is to develop new natural tools for 
the analysis and verification or computing and programming.
We are particularly interested in novel forms of calibrating 
computational resources by syntactic methods.
In \cite{LeivantM18} we already show that, by using finite structures
as focal concept, one obtains an abstract characterization
of primitive recursive mathematics based on the concept of loop variant,
familiar from program verification.

The remaining of the paper is organized as follows.
In \S 2 we lay out our fundamental concepts and notations.
Our theory \bxbf{FS} of finite structures is presented in \S 3, 
followed by its programming language counterpart \bxbf{ST} in \S 4.
We then prove in \S 5 an abstract generalization of Kleene's classical
theorem about the existential definability of computability, and
use it in \S 6 to show that \bxbf{FS} is mutually interpretable
in Peano Arithmetic.


\medskip
\bigskip

\Section{Finite structures}

\subsection{Finite functions}
Our basic notion is the {\em finite partial} structure,
in which function-identifiers are interpreted as partial-functions.
For example, we construe
binary strings as structures over the
vocabulary with a constant $\tte$ and unary function identifiers
\ttzero\ and \ttone.  E.g., \bxtt{011}
is taken to be the four element structure
$$
\tte \,
        \circ \ara{\ttscript{0}}
        \circ \ara{\ttscript{1}}
        \circ \ara{\ttscript{1}}
        \circ \zero
$$
Here \ttzero\ is interpreted as the partial-function defined only
for the leftmost atom, and \ttone\ as the partial-function
defined only for the second and third.
As in Gurevich's ASM, we posit our finite structures to live within a 
denumerable set $A$ of {\em atoms,}
i.e.\ unspecified and unstructured objects.
To accommodate non-denoting terms
we extend $A$ to a flat domain $A_\bot$, where in addition to the atoms we posit
a fresh object $\bot$, intended to denote ``undefined."  
The elements of $A$ are the {\em standard} elements of $A_\bot$.\fn{We 
refer neither to boolean values not to
``background structures" \cite{BlassG07} on top of atoms.}

By an {\em $A$-function} we mean a finite $k$-ary partial-function
over $A$, where $k \geq 0$; thus, the nullary $A$-functions are the 
atoms.  We identify an A-function $F$ with
the total function
$\;\; \tilde{F}:\; A_\bot^k\, \ra \, A_\bot \;$ satisfying
\begin{equation}\label{eq:strict}
\begin{array}{rcl}
\tilde{F}(a_1, \ldots, a_k) 
	& = & \min(F(a_1, \ldots, a_k),\, \bot)\\
	& = & \text{if } F(a_1, \ldots, a_k) = b \in A \; \text{ then } b \;
			\text{ else } \bot
\end{array} \end{equation}
Note that $\tilde{F}$ is necessarily strict, i.e.\
its output is $\bot$ whenever one of its inputs is.
For each $k$ we let $\emptyset_k$ be the empty $k$-ary $A$-function, i.e.\
the one that returns $\bot$ for every input.  
When in no danger of confusion we write $\emptyset$ for $\emptyset_k$.
The {\em scope} of $F$ is the set of atoms occurring in its entries.

Function partiality provides a natural representation of finite relations
over $A$ by partial functions, without recourse to booleans.
We represent a finite $k$-ary relation $R$ over $A$ ($k >0$)
by $k$-ary $A$-functions whose domain is $R$, notably
the function
$$
\grx_R(a_1, \ldots, a_k) \; = \; \text{ if } R(a_1, \ldots, a_k) \text{ then } 
	a_1 \text{ else } \bot
$$

\subsection{$A$-structures over a vocabulary}

A {\em vocabulary} is a finite set $V$
of function-identifiers, with each $\bff\in V$ assigned an {\em arity}
$\ar(\bff) \geq 0$. We refer to nullary function-identifiers as {\em tokens}
and to unary ones as {\em pointers.}
For the moment we might think of these identifiers as reserved names, 
but they can be construed just the same as (permanently free) variables.

An {\em entry} of an $A$-function $F$ is 
a tuple $\lng a_1 \ldots a_k ,b \rng$ 
where $b = F(a_1,\ldots, a_k) \neq \bot$.
\zero\grs\ is an {\em $A$-structure over $V$} if it
is a mapping that to each $\bff\,^k \in V$,
assigns a $k$-ary $A$-function $\grs(\bff)$,
said to be a {\em component} of \grs.
The {\em scope} of \grs\ is the union of the scopes of its components.

If \grs\ is an $A$-structure over $V$, and \grt\ an $A$-structure over $W \supseteq V$,
then \grt\ is an {\em expansion} of \grs\ (to $W$), and \grs\ a {\em reduct}
of $\grt$ (to $V$), if the two structures have
the same interpretation for identifiers in $V$ (the scope of \grt\ may
be strictly larger than that of \grs).
For $A$-structures \grs\ and \grt\ over the same vocabulary,
we say that \grs\ is a {\em substructure} of \grt\
if every entry of \grs\ is an entry of \grt.

A finite collection of structures is itself a structure:
if $\grs_i$ is a structure over $V_i$ ($i \in I$),
then the collection is
the union $\grs = \sbcup_{i\in I} \grs_i$  over
the disjoint union of the vocabularies.

\medskip

Given a vocabulary $V$, the set $\bxbf{Tm}_V$ of {\em $V$-terms} is 
generated by
\bi
\li $\bgrw \in \bxbf{Tm}_V$.
\li If $\bff\,^k\in V$,
and $\bft_1, \ldots , \bft_k \in \bxbf{Tm}_V$ then
$\bff\bft_1 \cdots \bft_k \, \in \bxbf{Tm}_V$
\ei
Note that we write function application in
formal terms without parentheses and commas, as in $\bff xy$ or $\bff \vec{x}$.
We optionally superscript function identifiers by their arity,
and implicitly posit that the arity of a function matches the number of
arguments displayed; thus writing $f^k\vec{a}$ assumes that
$\vec{a}$ is a vector of length $k$,
and $f\vec{a}$ (with no superscript) that 
the vector $\vec{a}$ is as long as $f$'s arity.
A term is {\em standard} if $\bfgrw$ does not occur in it.

\medskip

Given an $A$-structure \grs\ over $V$,  
the {\em value} of a $V$-term \bft\ in $\grs$,
denoted $\grs(\bft)$, is obtained by recurrence on \bft: 

\be
\li $\grs(\bfgrw) = \bot$
\li For $\bff\,^k \in V$,
$\grs(\bff \bft_1 \cdots \bft_k) =
	\grs(\bff)(\grs(\bft_1), \ldots , \;
			\grs(\bft_k))$
\ee


An atom $a \in A$ is {\em $V$-accessible} in \grs\
if $a= \grs(\bft)$ for some $\bft\in \bxbf{Tm}_V$.
An A-structure \grs\ over $V$ is {\em accessible} if every atom in the 
scope of \grs\ is $V$-accessible.
For example, if $V$ is without tokens
then no atom of a $V$-structure can be accessible.
If every atom in the scope of an accessible structure \grs\ is the value
of a {\em unique} $V$-term we say that \grs\ is {\em free}.
Note that an accessible A-structure \grs\ can fail to be free
even if all its components are injective:

\xfig{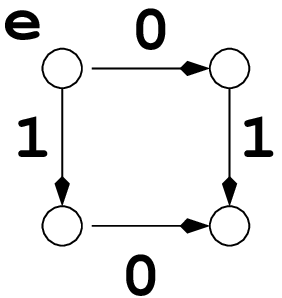}{16mm}

However, we have the straightforward observation:

\blem\label{lem:free}
An accessible A-structure \grs\ over $V$ is free 
iff there is a finite set $T$ of $V$-terms, closed under
taking sub-terms, such that the function $\grs:\; T \ra A$
is injective.
\elem

If \bfq\ is a standard $V$-term, and $T$ consists of the
sub-terms of \bfq, then we write $\calT(\bfq)$ for the resulting
{\em free} $A$-structure.  
It is often convenient to fix a reserved token,
say $\bullet$, to denote in each structure $\calT(\bfq)$
its root, i.e.\ the term \bfq\ as a whole.
%
For example, for nullary \ttz\ and \ttu, unary \tts, and binary \ttf,
the structures $\calT(\tts\tts\ttz)$,
$\calT(\ttf\ttz\ttu)$, 
$\calT(\ttf\tts\ttz\tts\ttz)$, 
$\calT(\ttf\ttz\tts\ttz)\;$ are, respectively,

\xfig{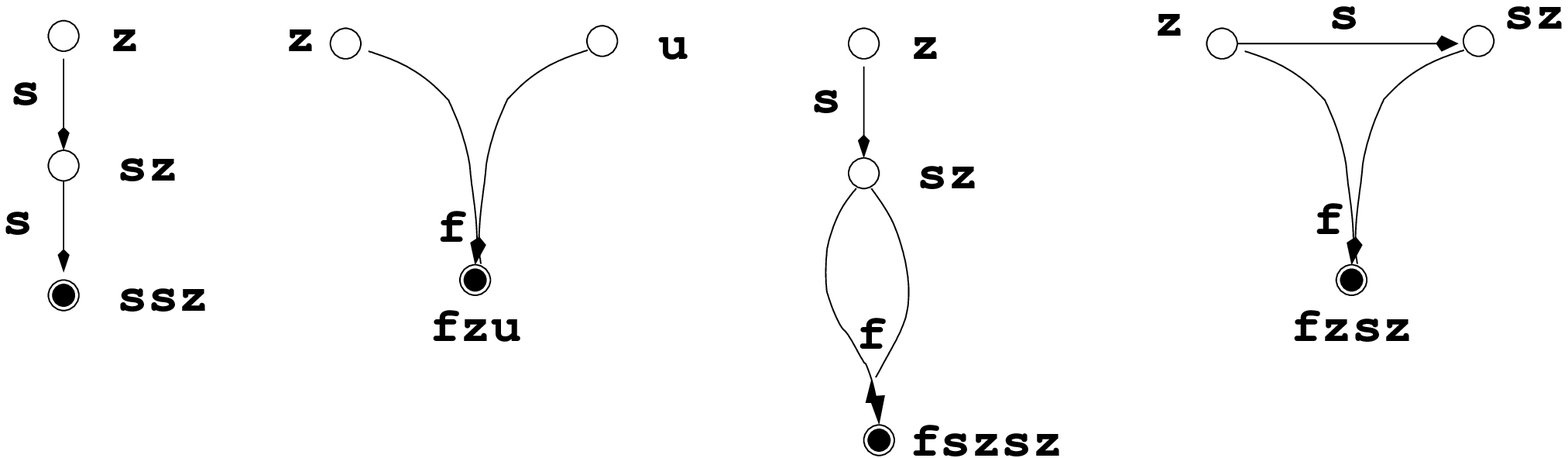}{30mm}

\subsection{A second-order language}

We wish to present a formal theory of finite structures,
that deals not only with one structure at a time, but with
structure transformation, in particular
by suitable imperative programs.
To that end, we generalize our discussion from a single
vocabulary $V$ to a vocabulary ``on demand."
We posit for each $k\geq 0$ a denumerable sets of {\em $k$-ary variables,}
intended to range over the $k$-ary $A$-functions.
The variables of arity $k \! = \! 0$ and $k \! > \! 0$ are dubbed {\em atomic}
and {\em functional}, respectively.
When a particular vocabulary $V = \bff_1^{r_1} \ldots \bff_k^{r_k}$
is of interest (with the exhibited ordering of its identifiers),
we write $\vec{g}^V$ (with the vocabulary's name as superscript)
for a vector of variables $g_1 \ldots g_k$, with $\ar(g_i)= r_i$.
We update our definitions of the set \bxbf{Tm} of {\em terms}, and
of their semantics in a structure \grs, by referring to function variables
rather than identifiers of a fixed vocabulary.

%

\medskip

An {\em equation} is a phrase of the form $\bft\deq \bfq$, where
$\bft, \bfq \in \bxbf{Tm}$.
The set $\bxbf{Fm}$ of {\em formulas} is generated inductively by
\be
\li Every equation is a formula;
\li $\bxbf{Fm}$ is closed under the propositional operators
	($\neg,\wedge,\vee$ and $\ra$).
\li If \grf\ is a formula, then so are 
$\forall f^r \; \grf$ and $\exists f^r \; \grf$ ($r \geq 0$).
\ee

Given an $A$-structure \grs\ for a set $W$ of variables,
a variable $f^k$, and $k$-ary $A$-function $F$,
we write $\{f:=F\}\grs$ for the structure $\grs'$ which is identical to \grs,
except that $\grs'(f) = F$.
Given a formula \grf\ and a structure \grs,
where all variables free in \grf\ are in the domain of \grs,
the {\em truth} of \grf\ in \grs,
denoted $\grs \models \grf$, is defined by recurrence on formulas,
as follows.

\begin{equation}\label{eq:tarski-Vfml}
\begin{array}{rclcl}
\grs & \models & \bft \deq \bfq & \bxsc{iff} &
	\grs(\bft) = \grs(\bfq) \\[1mm]
\grs & \models & \neg\grf & \bxsc{iff} &
	\grs\; \not\models \; \grf \\[1mm]
&&&& \minusthreecm \text{Similarly for other connectives} \\[1mm]
\grs & \models & \forall f^k \; \grf & \bxsc{iff} & 
	 \{f \la F\}\, \grs  \; \models \; \grf \quad \bxrm{for all finite }
		F: \; A^k \pa A \\[1mm]
\grs & \models & \exists f^k \; \grf & \bxsc{iff} & 
	 \{f \la F\}\, \grs  \; \models \; \grf \quad \bxrm{for some finite }
		F: \; A^k \pa A
\end{array}\end{equation}
When $\grs \models \grf$ we also say that \grs\ {\em verifies} \grf.
Clearly, $\grs \models \grf$ depends only on
$\grs(f)$ for variables occurring free in \grf. 

Recall from (\ref{eq:strict}) 
that we identify a finite $F: \; A^k \pa A$
with its strict extension $\tilde{F}: \; A_\bot^k \ra A_\bot$,
so the quantifiers range over all strict $\bot$-valued functions,
which for the nullary case $k \seq 0$ means that atomic variables
may take any value in $A_\bot$. 
This definition departs from Tarski's semantics in
that quantification ranges over {\em all} $A$-functions, regardless of the
scope of \grs.

\subsection{First-order formulas}

We call quantification over atomic variables {\em first-order},
and quantification over functional variables {\em second-order.}
A formula is {\em first-order} if all its quantifiers are first-order.

We use notation conventions for some important first-order formulas:
\bi
\li $\vec{\bft} \in f^k$ for $f\vec{\bft} \not\deq \bgrw$.
We overload this convention, and write
$\vec{\bft} \in f^1 $ to mean
$\bft_1 , \ldots , \bft_k \in f$,
where $\vec{\bft}= \bft_1 \ldots \bft_k$.

Dually, $\bft \in f^k$ abbreviates  \quad
$ \sbvee_{i+j=k-1} \;\;
\exists\, \vec{v}^i, \vec{w}^j\; \vec{v}\,\bft \vec{w} \in f$

\li\zero\\[-8mm]
\begin{equation}\label{eq:scope}
\begin{array}{rcll}
\bxrm{Scope}_V[x] & \equiv & \;\; \sbvee_{f^k \in V}
	\quad (\,x \in f\,) \vee (\,\exists \vec{v} \; x \deq f\vec{v}\,)
\end{array}
\end{equation}
states that the atom denoted by $x$ is in the scope of \grs.
Thus
$$
\grs \models \forall x \; \bxrm{Scope}_V[x] \sra \grf[x]
$$
means that \grf\ is true ``within" \grs, which 
is not the same as $\forall x \; \grf[x]$.
For example, $\forall x \; x \not\deq \bgrw$ is true for all $A$-structures,
whereas $\forall x \; \bxrm{Scope}_V[x] \sra x\not\deq \bgrw$
is false for all $A$-structures.  
\li Let $V=\{\ttz^0,\tts^1\}$.
The $V$-structures that model the following formula are precisely
the structures $\calT(\tts^{i}\ttz)$ for the natural numbers, described 
in \S 2.2.\fn{We disregard here the token $\bullet$}
\begin{equation}\label{eq:fmlForN}
\begin{array}{ll}
\grn \quad \equiv \quad 
	& (\;\forall x,y\;\; \tts x \!\deq\! \tts y \; \wedge \;
		x \!\not\deq\! \bgrw \; \wedge\; y\! \not\deq \! \bgrw 
		\;\; \ra \;\; x\!\deq\!y\;) \\[1mm] 
	& \quad  \wedge \;\; (\forall x\; \;
		\tts x \!\deq\! \ttz \;  \vee \; \exists y \; x\!\deq\!\tts y \;)
\end{array}
\end{equation}
\ei

The following observation is a variant of a basic result of Finite Model Theory.
We shall not use it in this paper.  

\bprop\label{prop:FO-decidable}
Let \grf\ be a first-order formula.
The following problem is decidable in polynomial time (in the
size of the input structure presentation).  We assume
that $A$-structure are given as tables.
\bi
\li {\sl Given an A-structure \grs, is \grf\ true for \grs.}
\ei
\eprop
\prf
The proof is by induction on \grf.
Absent second-order quantifiers,
the only case of interest is when the main operator of \grf\
is an atomic quantifier, say $\forall u$.  The input's format
provides direct scanning of all entries that are in the scope of \grs,
and the denotations of the A-functions all yield $\bot$
for all other entries.  From this the calculation of
the truth in $\grs$  of $\forall u \grf$ is immediate.
\qed

%

%

\Section{A theory of finite structures}

\subsection{On axiomatizing finiteness}

Since our $A$-structure are built of functions,
we refer to a generative process that extends functions rather than sets
ore relations.  The set $\gothF^k$ of $k$-ary $A$-functions is
generated by:
\bi
\li {\sl The empty $k$-ary function $\bemptyset_k$ is in $\gothF^k$;}
\li {\sl If $F \in \gothF^k$,
$v \in \vec{u} \in A$, and $F(\vec{u}) = \bot$, then
extending $F$ with an entry 
$F\vec{u}=v$ yields an $A$-function in $\gothF_k$.}
\ei

\medskip

From this inductive definition
we obtain an Induction Schema for $A$-functions.
Using the abbreviations
\begin{equation}\label{eq:induction-abbreviations}
\begin{array}{rcl}
\grf[\emptyset] & :\equiv & \forall g \; (\forall \vec{w} \; g\vec{w}=\bgrw)
				\;\; \ra \; \grf[g]\\[2mm] 
\text{and} \quad \grf[\{\vec{u}\mapsto v\}\,f]
		& :\equiv &
		\forall g \; ( \; g\vec{u} \deq v \;\;
	\wedge \;\; \forall \vec{w}\not\deq \vec{u} \;\;\; g\vec{w} \deq f\vec{w} \; )
		\;\; \ra \;\; \grf[g]
\end{array}
\end{equation}
Induction for a formula $\grf[f^k]$
(with a distinguished function-variable $f$) reads
\begin{equation}\label{eq:functionInduction}
\begin{array}{l}
(\forall f^k,\vec{u},v \;\;\; \grf[f] \; \wedge \; (f\vec{u} \! \deq \! \bgrw)
	\;\ra \; \grf[\{\vec{u}\mapsto v\}\,f] \; ) \\
\qquad \ra \;\; \grf[\emptyset] \; \ra \; \forall f \;\grf[f]
\end{array}\end{equation}


	
Since the components of $A$-structures are $A$-functions, without
constraints that relate them,
there is no need to articulate a separate induction principle for 
$A$-structures.
Indeed, every $A$-structure \grs\ for a vocabulary $V=\{f_1, \ldots,f_k\}$
is obtained by first generating the entries of $\grs(f_1)$,
then those for $\grs(f_2)$, and so on.
Note that this would no longer be the case for an inductive
definition of the class $\gothA$ of {\em accessible} structures, 
whose components must be generated in tandem:\fn{We shall discuss 
the class $\gothA$ in detail elsewhere.}
\bi

\li {\sl If $\grs = (F_1 \ldots F_k)$ is in $\gothA$,
\ul{$\vec{a}$ is {\em in the scope of \grs,}}
$(\grs\bff)(\vec{a}) = \bot$,
and $b \in A$,\\
then extending \grs\ with an entry
$F_i\vec{a}=b$ yields a structure in $\gothA$.}
\ei

\subsection{The theory \bxbf{FS}}

Our axiomatic theory \bxbf{FS} for $A$-functions has free and bound variables
for {\em atoms} and free and bound variables for {\em functions} of
arbitrary positive arity.  We use $u,v,\ldots$ as syntactic parameters
for atomic variables, and $f,g,h, \ldots$ as syntactic parameters
for functional variables, optionally superscripted with their arity
when convenient.

The axiom schemas are the universal closures 
of the following templates, for all arities $i,k$,
terms \bft, and formulas \grf.
For a function variable $f$ and variables $\vec{u}$ 
we write $\vec{u} \in f$
for $f\vec{u} \not\deq \bgrw$.

\bd
\li[\bxbf{Strictness}] \qquad\qquad
	$fu_1 \cdots u_k \deq \bgrw \;\; 
		\ra \;\; \sbvee_i u_i \! \deq \! \bgrw$

\li[\bxbf{Infinity}]\qquad \qquad
$\exists \vec{u} \; f^k\vec{u} \deq \bgrw$ 

Note that this states that $A$ is infinite (unbounded), and has no bearing
on the finiteness of $A$-functions.

\li[\bxbf{Empty-function}] \qquad\qquad
	$\exists g^k \; \forall \vec{u}. \; g\vec{u} \; \deq \; \bgrw$.

\li[\bxbf{Extension}] \qquad\qquad
	$\exists g \;
	g\vec{u} \deq v \;\; \wedge \;\; 
		\forall \vec{w}\not\deq \vec{u} \;\;\; g\vec{w} \deq f\vec{w}$

\li[\bxbf{Explicit-definition}] \qquad
$
\exists g\, \forall \vec{u} \;\;
(\;\vec{u} \in f \;\; \wedge \;\; g\vec{u}\deq\bft\; )
\; \vee \; 
(\; \vec{u} \not\in f  \;\; \wedge \; \; g\vec{u} \deq \bgrw \;)
$ \\[1mm]
This schema combines Zermelo's Separation Schema
with an Explicit Definition principle:
$g$ is defined by the term \bft, for arguments in the domain of $f$.

\li[\bxbf{f-Induction}] \qquad Recalling (\ref{eq:functionInduction})
above:
$$
\begin{array}{l}
( \;\forall f,\vec{u},v \;\;\;
\grf[f] \;\; \wedge \; \; (f\vec{u}\! \deq \! \bgrw)
        \;\;\ra \;\; \grf[\{\vec{u}\mapsto v\}\,f]\;) \\[1mm]
\qquad \ra \;\; \grf[\emptyset] \; \ra \; \forall f \;\grf[f]
\end{array}
$$

%
%

\ed
For each of the schemas [Empty-function], [Function-extension],
and [Explicit-definition], the function $g$ asserted to exist is
trivially unique, and so adding identifiers
for these functions is conservative over \bxbf{FS}. 
We write $\bemptyset_k$, $\{\vec{u}\mapsto v\}\,f$ 
and $\grl \vec{u}\in f.\, \bft$,
respectively, for these functions (one for every arity $k \geq 1$).

The relevant axiom-schemas above can be rephrased
using these three constructs as primitives:

\bd
\li[\bxbf{Empty-function}] \qquad\qquad
	$\emptyset_ku_1 \cdots u_k \deq \bgrw$

\li[\bxbf{Extension}] \qquad\qquad
	$(\{\vec{u}\mapsto v\}\,f)\, \vec{u} \deq v \;\; \wedge \;\; 
		\forall \vec{w}\not\deq \vec{u} \;\;\; 
		(\{\vec{u}\mapsto v\}\,f)\, \vec{w} \deq f\vec{w}$

\li[\bxbf{Explicit-definition}] \qquad
$
\forall \vec{u} \;\;
(\;\vec{u} \in f \;\; \wedge \;\; (\grl \vec{v} \in f)\, \vec{u}\deq\bft\; )$\\
\zero\fivecm
$ \vee \; 
(\; \vec{u} \not\in f  \;\; \wedge \; \; (\,\grl \vec{v} \in f)\,\vec{u} \deq \bgrw \;)
$
\ed

Also note the contra-positive form of f-Induction:
\begin{equation}\begin{array}{l}
(\forall f,\vec{u},v \;\;
	 \grq[\{\vec{u} \mapsto v\} f] \; \ra \; \grq[f])
                \;\; \ra \;\; (\grq[h] \ra \grq[\emptyset])
\end{array}\end{equation}

\subsection{Some derived schemas}

\bd
\li[\bxbf{Union}]
	$\exists g \; \forall \vec{u} \; \vec{u} \in g \; \lra \;
		(\, \vec{u} \in f_1 \; \vee \; \vec{u} \in f_2\, )$.\\	
The proof is by f-Induction on $f_2$ and Extension.

\li[\bxbf{Composition}]\\[-8mm]
\begin{equation}\label{eq:composition}
\exists h^k \; \forall \vec{x} \;\; h\vec{x} \deq f^i(g_1\vec{x}) \cdots (g_i\vec{x}) 
\end{equation}
This follows from Explicit-definition and Union.

\li[\bxbf{Branching}] \qquad
	$\exists h^k \; \; \forall \vec{x}
	\zero\qquad  ( \;f\vec{x}\not\deq \bgrw \; \wedge \; h\vec{x}\deq f\vec{x}\; )
		\;\; \vee \;\;  
	( \;f\vec{x}\deq\bgrw \; \wedge \;  h\vec{x} \deq g\vec{x}\; )$

The universal closure with respect to $f,g$
is proved by inductions on $f$ and $g$, using Extension.

\li[\bxbf{Contraction}] \\[-8mm]
\begin{equation}\label{eq:contraction}
\exists g \;\;
 g\vec{u} \deq \bgrw \; \wedge \;
                \forall \vec{w}\not\deq \vec{u} \;\; g\vec{w} \deq f\vec{w}
\end{equation}
This is the dual of Extension.
The proof of (ref{eq:contraction}) 
is by induction on $f$ for the universal closure
of (\ref{eq:contraction}) with respect to $\vec{u}$.

\li[\bxbf{Function pairing}]\\[-8mm]

\begin{equation}\label{eq:function-pairing}
\forall f^k, g^k \; \exists h^{k+1}\; \exists a,b\;
	\forall \vec{x} \; f\vec{x} \deq ha\vec{x} 
		\; \wedge \; g\vec{x} \deq hb\vec{x}
\end{equation} 
Much of the expressive and proof theoretic power of 
arithmetic is due to the
representation of finite sequences and finite sets of numbers by  numbers.
The Function-pairing schema provides
a representation of two $k$-ary $A$-functions by a single
$(k \splus 1)$-ary $A$-function.  Namely, $f,g$ are ``tagged" withing
$h$ by the tags $a,b$ respectively.
In other words, writing $h_u$ for $\lambda \vec{x}\, hu\vec{x}$,
we have $f = h_a$ and $g= h_b$.
(We might require, in addition, that $hu\vec{x} = \bot$ for all atoms
$u \neq a,b$, but this is inessential if we include $a,b$ explicitly in the
representation.)

\li[\bxbf{Atomic-choice}] 
A more interesting form of tagging is provided by the following principles
of {\em Choice}.  
\begin{equation}
\zero\qquad
(\;\forall {x} \in f \; \exists y \; \grf[\vec{x},y] \;)
	\; \ra \; 
	\exists g\; \forall \vec{x}\! \in \! f \;\;  \grf[\vec{x},g\vec{x}]
\end{equation}
This is analogous to \cite[Lemma 2(e)]{Parsons70},
and is straightforward by induction on $f$. 
Suppose the schema holds for $f$, yielding the function $g$.
To show the schema true for $f' = \{\vec{u}\mapsto v\}f$ suppose
it satisfies the premise
\begin{equation}\label{eq:choice-premise} 
\forall \vec{x}\in f \; \exists y \; \grf[\vec{x},y] 
\end{equation}
Then $f$ satisfies (\ref{eq:choice-premise}) as well, 
yielding a $g$ for the conclusion.
Also, by (\ref{eq:choice-premise})  there is
an atom $y$ such that $\grf[\vec{u},y]$, and so the conclusion is satisfied by
$\{\vec{u}\mapsto y \} g$ in place of $g$.

\li[\bxbf{Function-choice}] \\[-8mm]
$$ \zero\qquad (\forall \vec{x} \in f \;\; \exists g \; \grf[\vec{x},g])
        \; \ra \;
        \exists h\; \forall \vec{x} \! \in \! f \;  \grf[\vec{x},h_{\vec{x}}]
$$
where $\grf[\vec{x},h_{\vec{x}}]$ \quad abbreviates \quad
$ \forall j \; (\;\;(\forall \vec{y} \; j\vec{y} \deq h\vec{x}\vec{y}) \; \ra \;
		\grf[\vec{x},j]\; ). $
Note that Atomic-choice is a special case of Function-choice, with
$g$ nullary.

Note that the bounding condition in the choice schemas above is essential:
even the simplest case
\begin{equation}\label{eq:general-choice}
(\forall x^0 \exists y^0 \grf[x,y]) \;\; \ra \;\;
	\exists f^1 \forall x \; \grf[x,fx] 
\end{equation}
is false already for $\grf \equiv x\deq y$, since the identity function
over $A$ is not finite.
\ed

\Section{Imperative programs over $A$-structures}

We define a variant of Gurevich's abstract state machines (ASMs)
\cite{Borger02,Gurevich93e,Gurevich01}, for the transformation
of $A$-structures.  Transducer-programs define mappings between structures,
which are akin to the mappings underlying Fraenkel's Replacement
Axiom, and more generally the proper classes of Bernays-G\"{o}del set theory.
Namely, these mappings are not $A$-functions, and are referred to
in the theory \bxbf{FS} via the formulas that define them.


\bigskip

\subsection{Structure revisions}

Our {\em structure transformation} programming language,
\bxbf{ST}, is designed to be a Turing-complete computing system for the
transformation of finite partial-structures, using the simplest possible
building blocks while maintaining expressiveness.
An \bxbf{ST}-program takes an $A$-structure as input, and successively
applies basic structure revisions to it. The process may terminate
with a final $A$-structure when no further revision is called for.
For example, addition over \dN\
might be computed by a program that takes as input a structure
representing two natural numbers $n_1$ and $n_2$, 
and grafting the second on top of the first.
Doubling a number might be performed by copying the input before grafting
the copy over the input,
or alternatively generating a new structure by repeatedly
extending it with two atoms while depleting the input. 

We start by defining the basic operations of \bxbf{ST}, to handle entries.
Each such operation maps a structure \grs\ into a structure $\grs'$
which is identical to \grs\ with the exceptions noted.

\be
\li An {\em extension} is a phrase
$\bff\, \bft_1 \cdots \bft_k := \bfq$ where
the $\bft_i$'s are standard terms and \bfq\ 
is a term.
The intent is that $\grs'$ is identical
to \grs, except that if
$\grs(\bff\, \bft_1 \cdots \bft_k) = \bot$ then
$\grs'(\bff\, \bft_1 \cdots \bft_k) = \grs(\bfq)$.

\li An {\em inception} is a phrase of the form 
$\bfc \ssDownarrow$, where \bfc\ is a token.
A common alternative notation is $\bfc := \bxbf{new}$. 
The intent is that $\grs'$ is identical to \grs,
except that if $\grs(\bfc) = \bot$, then $\grs'(\bfc)$ is
an atom not in the scope of \grs.
A more general form of inception, with a fresh atom assigned
to a term \bft\ is obtained as the composition
$$
\bfc \sDownarrow; \;\; \bff\gra_1\cdots\gra_k := \bfc; \;\; \bfc := \bgrw
$$
where \bfc\ is a reserved token.

We allow extensions and inceptions to refer to an identifier \bff\ not in the
vocabulary of \grs, in which case the vocabulary
of $\grs'$ extends that of \grs, and we posit that $\grs(\bff)= \emptyset$.


\li A {\em contraction,} the dual of an extension,
is a phrase of the form
$\bff\gra_1 \cdots \gra_k \suparrow$.
The intent is that $\grs'$ is identical to \grs,
except that $\grs'(\bff\gra_1 \cdots \gra_k) = \bot$.

\li A {\em deletion,} the dual of an inception, is a phrase of the form
$\bfc \ssUparrow$, where $\bfc$ is a token.
The intent is that $\grs'$ is obtained by
removing $\grs(\bfc)$.  That is, $\grs'$ is identical
to \grs, except that for all $A$-functions $f^k$ present
and all $a_1 \ldots a_k \in A$,
if $\grs(f)\,\vec{a}=\grs(\bfc)$,
then $\grs'(f)\vec{a} = \bot$.

Note that a deletion cannot be obtained via the composition of
contractions, because $\grs(c)$ might be reached by $A$-functions
of \grs\ from atoms that are not accessible, in which case 
the atom $\grs(\bfc)$ cannot be eliminated 
from the scope of $\grs'$ by contractions alone.
\ee

%
We refer to extensions, contractions, inceptions and
deletions as {\em revisions}.
Extensions and inceptions are then
{\em constructive revisions,} whereas contractions and deletions are the
{\em destructive revisions.}


An extension and a contraction
can be combined into an
{\em assignment}, i.e.\ a phrase of the form $\bff\vec{\bft} := \bfq$.
This can be viewed as an abbreviation, with \ttb\ a fresh token, 
for the composition of four revisions:
$$
\ttb \sdownarrow \bfq; \;\;
\bff\vec{\bft}\uparrow; \;\;
\bff\vec{\bft} \sdownarrow \ttb; \;\;
\ttb\suparrow
$$
(Note that the atom denoted by \bfq\ (when defined) is being
memorized by \ttb, since \bfq\ may become inaccessible due to the
contraction $\bff\vec{\bft}\uparrow$.)
Although assignments are common and useful,
we prefer contractions and extensions as basic constructs, for two reasons.
First, these constructs are truly elemental.
More concretely, the distinction between constructive and
destructive revisions plays a role in implicit characterizations of
computational complexity classes, as for example in
\cite{LeivantM18}.




\subsection{\bxbf{ST} programs}

Our programming language \bxbf{ST} for structure transformation 
consists of guarded iterative programs using revisions as basic operations.
Define a {\em guard} to be a quantifier-free formula.\fn{Taking for 
guards arbitrary {\em first-order} formulas would not make a difference
anywhere.}
The {\em programs of \bxbf{ST}}
are generated inductively as follows.

\be
\li A structure-revision is a program.
\li If $P$ and $Q$ are programs 
and $G$ is a guard, then  $P;\,Q$,
$\;\bxbf{if}\,[G]\,\{P\}\,\{Q\}\;$ and
$\; \bxbf{do}\;[G]\,\{P\}\;$ are programs.
\ee


For a program $P$ we define
the binary {\em yield relation} $\rA_P$ between 
structures by recurrence on $P$.
For $P$ a revision the definition follows the intended semantics
described informally above.  The cases for composition, branching,
and iteration, are straightforward as usual.

Let $\grF:\; \gothC \pa \gothC'$ be a partial-mapping from a class
\gothC\ of A-structures to a class $\gothC'$ of A-structures.
A program $P$ {\em computes \grF} if for every
$\grs \in \gothC$, $\grs \rA_P \grt$ for some expansion \grt\
of $\grF(\grs)$.
Note that the vocabulary of the output structure
need not be related to the input vocabulary.



\subsection{Turing completeness}
Guarded iterative programs are well known to be sound and complete 
for Turing computability, and proofs of the Turing completeness of abstract
state machines have been given before (see for example
\cite[\S 3.1]{Gurevich00}).  To dispel any concern that
those proofs need more than finite structures and our
simple revision operations, we outline a proof here.

\begin{thm}\label{thm:turing-completeness}
Let $M$ be a Turing transducer 
computing a partial-function $f_M: \; \grS^* \sra \grS^*$.
There is an \bxbf{ST}-program $P_M$ that, for every $w \in \grS^*$,
transforms $\calT(w)$ to $\calT(f_M(w))$.
\end{thm}
\prf
Suppose $M$ uses an extended alphabet $\grG \supset \grS$, 
set of states $Q$, start state
\tts, print state \ttp, and transition function \grd.
Recall that for $w = \grg_1 \cdots \grg_k \in \grG^*$
we write $\calT(w)$ for
the structure 
$\tte \circ \ara{\gamma_1} \circ \cdot\cdot\cdot \circ \ara{\gamma_k} \circ$.
  
Define $V_M$ to be the vocabulary with
$\tte$, \ttc\ and each state in $Q$ as tokens; and with
\ttr\ and each symbol in \grG\ as pointers.
The intent is that a configuration 
$(q, \grs_1 \cdots \underline{\sigma}_i \, \cdots \grs_k)$ (i.e.\ with
$\underline{\sigma}_i$ cursored)
be represented by the $V_M$-structure
$$ \begin{array}{rccccccccc}
& \circ & \!\!\!\! \ara{\sigma_1} \,\,\, \circ & \cdot\cdot\cdot &
	& \circ & \!\!\!\! \ara{\sigma_i} \,\, \circ & \cdot\cdot\cdot &
		\circ \; \ara{\sigma_k} \; \circ\\[-1mm] \;
& \tte,\ttq & & && \ttc &
\end{array}$$
All remaining tokens are undefined.

We define the program $P_M$ to implement the following phases:
\be
\li Convert the input structure into the structure for the initial
configuration, and initialize \ttc\ to the initial input element.
Use a loop to initialize a fresh pointer \ttr\ to be the destructor function 
for the input string, to be used for backwards movements of the cursor.
\li Main loop: configurations are revised as called for by \grd.
The loop's guard is \ttp\ (the ``print" state) being undefined.
\li Convert the final configuration into the output.
\ee 
\qed

\Section{Computability implies existential definability}

\subsection{Expressing relation iteration}

Suppose $\grf[\vec{f}\,^V,\vec{g}\,^V]$ is a formula, 
with variable-vectors $\vec{f},\vec{g}$, both for the 
vocabulary $V$.
Define the relation $[\grf]^*[\vec{f}\,^V,\vec{g}\,^V]$ as the least fixpoint 
of the closure conditions:
\bi
\li $[\grf]^*[\vec{f}\,^V,\vec{f}\,^V]$ for all $\vec{f}\,^V$.
\li If $\;\grf[\vec{f}\,^V,\vec{g}\,^V]\;$ and
$\; [\grf]^*[\vec{g}\,^V,\vec{h}\,^V]\;$
then
$\;[\grf]^*[\vec{g}\,^V,\vec{h}\,^V]$.
\ei

\begin{thm}\label{thm:fixpoint}
For every formula $\grf[\vec{f}\,^V,\vec{g}\,^V]$ there is a formula
$\grf^*[\vec{f}\,^V,\vec{g}\,^V]$ that defines the relation
$[\grf]^*[\vec{f}\,^V,\vec{g}\,^V]$.
\end{thm}
\prf
By definition, $[\grf]^*[\vec{f}\,^V,\vec{g}\,^V]$ just in case there are
$\vec{h}_0^V, \ldots ,\,\vec{h}_k^V$ ($k \geq 0$) such that
$$
\vec{h}_0 = \vec{f}, \qquad
\grf[\vec{h}_i^V,\vec{h}_{i+1}^V] \quad (i<k),\qquad \text{and} \quad
\vec{h}_k = \vec{g}
$$

The function-vectors $\vec{h}_i$ can be bundled jointly 
into a single vector $\vec{h}$, using
a vocabulary $\hat{V}$ with a fresh variable $\bar{f}^{k+1}$ for each $f$ in $V$. We also
refer to two fresh auxiliary variables $z^0$ and $s^1$, and let the 
bundled vector $\vec{h}$ be defined by
$$
\vec{h}\,^{\hat{V}}(s^{[i]}z)\vec{x}) = h_i(\vec{x})
$$
For a term \bft\ let $\vec{h}_{\bf t} = \vec{\ell}$
abbreviate $\forall \vec{u} \; \vec{\ell}\vec{u} \deq \vec{h}\bft\vec{u}$.

Now define
\begin{equation}\label{eq:bundle}
\begin{array}{l}
\grf^*[\vec{f}\,^V,\vec{g}\,^V] \quad \equiv \\[1mm]
\zero \quad \begin{array}[t]{l}
\exists z^0,s^1
 ( z \not\deq \bgrw \; \wedge \;  
	\forall x^0,y^0 \; (sx \not\deq z) \wedge (sx\seq sy \sra x\seq y)\\[1mm]
\wedge \;\; 
		\exists \vec{h}\,^{\hat{V}} \;\; \vec{h}_z = \vec{f} \\[1mm] 
	\quad\quad \wedge \;\; \forall x \; ( \; sx \not\deq \bgrw \; \ra \;	
		(\forall \vec{\ell}\,^V \; \vec{m}\,^V \;
			\vec{\ell} = \vec{h}_x
			\; \wedge \; \vec{m} = \vec{h}_{sx}
			\; \ra \; \grf[\vec{\ell},\vec{m}]\;) \\[1mm]
	\quad\qquad\qquad  \wedge \;\; ( \; sx \deq \bgrw  \; \wedge x \not\deq \bgrw 
			\; \ra \; \vec{h}_x = \vec{g}\;))
\end{array}
\end{array}
\end{equation}
\qed

\noindent{\bxbf{Example.}
Let $V$ be a vocabulary with identifiers of arity $\leq k$,
and $\vec{x},\vec{y}$ etc.\ be vectors of $k$ atomic variables.
Take for $\grf[g^1,h^1]$ the following formula (where 
$g$ and $h$ are construed as sets). 
$$
\forall x \; (\; hx\deq gx \;\vee \;\; 
	\sbvee_{f^{r} \in V} 
	\exists y_1 \ldots y_r \in g \; f\vec{y} \in h \;\; \wedge\;\;
			f\vec{y}\not\in g 
$$
That is, $h$ is identical to $g$ except that it may contain
additional atoms, all obtained by applying some $f \in V$
to elements of $g$.
Then the existential formula
$$
A_V[u]  \; \equiv \;\;
	\exists g^1 \; \grf^*[\emptyset,g] \;\wedge \; u \in g
$$
defines the set of $V$-accessible atoms.  Consequently,
$$
\forall u \; (\sbvee_{f \in V} \; u\in f)
	\;\;  \ra\; A_V[u]
$$
is true for a $V$-structure \grs\ iff \grs\ is $V$-accessible.

\subsection{Existential definability of computable relations between $A$-structures}

\begin{thm}\label{thm:compToSigma}
For every \bxbf{ST}-program $P$ over $V$-structures,
there is an existential formula $\grf_P[\vec{f}\,^V, \vec{g}\,^V]$
that holds iff $\vec{f} \rA_P \vec{g}$.
\end{thm}
\prf By induction on $P$. 

\bi
\li
If $P$ is a revision then $\grf_P$
is in fact first-order, and trivially defined.

\li
If $P$ is $Q;R$, then by IH there are existential formulas 
$\grf_Q[\vec{f},\vec{h}]$ and $\grf_R[\vec{h},\vec{f}]$ that 
hold just in case $\vec{f} \rA_Q \vec{h}$ and 
$\vec{h} \rA_Q \vec{g}$.  We thus define
$$
\grf_P [\vec{f},\vec{g}] \quad \equiv \qquad
	\exists \vec{h} \;\; \grf_Q[\vec{f},\vec{h}] \; \wedge \;
	\grf_R[\vec{h},\vec{f}]
$$
which is existential if $\grf_A$ and $\grf_R$ are.

\li If $P$ is  $\bxbf{if}[G]\{Q\}\{R\}$ we define
$$
\grf_P[\vec{f},\vec{g}] \quad \equiv \qquad
	(\;G[\vec{f}] \; \wedge\; \grf_Q[\vec{f},\vec{g}]\;)
	\;\; \vee \;\;
	(\;\neg G[\vec{f}] \; \wedge\; \grf_R[\vec{f},\vec{g}]\;)
$$
which is existential if $\grf_Q$ and $\grf_R$ are.
Note that $G$ here is first-order, so the negation is harmless.

\li If $P$ is  $\bxbf{do}[G]\{Q\}$,
let 
$$
\grq[\vec{h},\vec{j}] \quad \equiv \qquad
	G[\vec{h}] \; \wedge \; \grf_Q[\vec{h},\vec{j}]
$$
and define
$$
\grf_P[\vec{f},\vec{g}] \quad \equiv \qquad
	\grq^\star[\vec{f},\vec{g}] \; \wedge \; \neg\, G[\vec{g}]
$$
Here $\grq^\star$ is the existential formula 
defined in (\ref{eq:bundle}).\\[-1cm]
\ei
\qed

Theorem \ref{thm:compToSigma} defines a binary relation between
the initial and final configurations of an \bxbf{ST}-program.
Given a convention on which program variables
are to be considered inputs and which output, the formula $\grf_P$
can be modified to express the input-output relation computed
by the program. For instance, if $P$ uses
variables $f_1 \ldots f_k$, of which $f_1$ and $f_2$ are used for the inputs,
and $f_2$ and $f_3$ are used for the outputs,
then the I/O relation defined by $P$ is
\begin{equation}\label{eq:IO}
\bxit{IO}_P[g_1,g_2;g_2',g_3'] \quad \equiv \quad
\exists g_1', g_4' \ldots g_k' \;
	 \grf_P[g_1,g_2,\vec{\bemptyset};g_1', \ldots, g_k']
\end{equation}

%

%
%
%

\Section{Equipollence of \bxbf{FS} and \bxbf{PA}}

\subsection{Interpretation of arithmetic formulas in finite structures}

The intended model of \bxbf{FS} has no ``universe" in the traditional, 
Tarskian, sense. Using the structures $\calT(\tts^{[n]}\ttz)$ as the
target ``elements" falls short of the natural embedding,
where natural numbers are interpreted as
equivalence classes of such structures.
Note that we cannot take
one representative from each equivalence class because there is
no way to formally identify such representatives.
Consequently, we depart from the traditional definition of interpretations 
between languages and between formal theories,
(see e.g.\ \cite[\S 2.7]{Enderton2001}),
relax the requirement that the source universe be interpreted
by a definable subset of the target universe, 
and interpret the natural numbers instead by equivalence classes of
structures satisfying \grn\ (as defined in (\ref{eq:fmlForN})).

The representation of natural numbers by equivalence classes of structures
can now be formalized as follows. 
Take \bxbf{PA} to be based on logic {\em without} equality, that is with equality
considered a binary relation identifier rather than a logical constant,
which for \bxbf{PA} happens to be interpreted
as identity.  The point is that our interpretation of \bxbf{PA}
in \bxbf{FS} must now include a definition of the interpretation of 
equality, though not as the identity relation between the atoms of \bxbf{FS}, 
but rather as structure isomorphism.
The property of a unary function $f$ being an isomorphism
between structures over the vocabulary $z^0,s^1$
can be defined by
\begin{equation}\label{eq:isom} \begin{array}{l}
 \bxrm{Isom}[f;z,s;z',s'] \;\; \equiv \;\;
  	fz = z' \\[1mm]
 \qquad \wedge \;\;  
 	\forall a \;\; sa \neq \bgrw \;\; \ra \;\; (\,fsa = s'fa \neq \bgrw\, )
 \end{array} \end{equation}
Given that \grn\ (as defined in (\ref{eq:isom})) is assumed true for
$(z,s)$ and $(z',s')$, and
that $f$ must be strict, the condition
$\bxrm{Isom}[f;z,s;z',s']$ above implies that the 
function $f$ must be a bijection.
We now define the interpretation of the equality relation of \bxbf{PA}
as the relation
$$
\bxrm{Iso}[z,s;z',s'] \quad \equiv \quad
\exists f^1 \;\;\bxrm{Isom}[z,s;z's']
$$
between structures representing natural numbers.

We interpret the remaining non-logical constants of \bxbf{PA},
namely 0, successor, addition and multiplication,
via basic equations for them.
For each variable $x_i$ of \bxbf{PA} let $z_i^0$ and $s_i^1$ be
\bxbf{FS}-variables.  In any given context only finitely
many \dN-variables will be present, so the collection of all
corresponding \bxbf{FS}-variables will be finite as well.
Every equation of $\bxbf{PA}$ is equivalent to a formula involving
only equation of one of the following five forms:
\begin{equation}\label{eq:5equations}
x_i =0, \;\; x_i = x_j, \;\; x_i = sx_j, \;\; x_i=x_j+x_k, \;\; \text{and} \;\;
x_i = x_j\cdot x_k
\end{equation}
We define the following interpretation for such equations.
\bi
\li $(x_i = 0)^{\btu} \; \equiv \; 
	(z_i \neq \bgrw \wedge s_i = \bemptyset)$.

\li $(x_i = x_j)^{\btu} \; \equiv \;  
	\exists f^1 \; \bxrm{Iso}[f;z_i,s_i;z_j,s_j]$
\qquad where \bxrm{Iso} is as in (\ref{eq:isom}).

\li 
$ (x_i = sx_j)^{\btu} \; \equiv \;  
	\exists t^1 \;\;\bxrm{Suc}[s_j;t] \;\; \wedge \;\;
		\bxrm{Iso}[z_j,t;z_i,s_i] $\\[1mm]
where \quad $\bxrm{Suc}[s,t]$ \quad  is
\begin{equation}\begin{array}{l}
        \forall a \;\; a \neq \bgrw \qquad  \ra \\
\qquad  (sa \neq \bgrw \;\; \wedge \;\; \,ta = sa) \\
\qquad \;\; \vee \;\;
	(sa = \bgrw \;\wedge \; ta \neq \bgrw \; \wedge \; tta = \bgrw\,)
\end{array} \end{equation}

\li $(x_i=x_j+x_k)^{\btu} \; \equiv \; \exists\, z^0, t^1 \;\;
	\grf_A[z_j,s_j,z_k,s_k;z,t] \;\; \wedge \;\;
	\bxrm{Iso}[z,t;z_i,s_i]$

where $\grf_A$ is the existential formula defined by
(\ref{eq:IO}) for the \bxbf{ST}-program $A$ computing addition,

\li ($x_i=x_j\times x_k)^{\btu}$ is defined similarly, referring to the
\bxbf{ST}-program for multiplication.

\li Combining the previous cases, we easily define (by discourse-level
induction on terms) an interpretation $(\bft=\bfq)^{\btu}$
for all terms \bft,\bfq.
\ei

We let the mapping $\btu$ commute with the connectives:
$(\grf\wedge\grq)^{\btu}$ is $\grf^{\btu} \,\wedge\, \grq^{\btu}$,
etc.

For the quantifiers we let
\begin{equation}
\begin{array}{rcl}
(\forall x_i \; \grq)^{\btu} 
	& \; \equiv \; & \forall z_i, s_i \;\; \grn[z_i,s_i]
		\;\; \ra \;\; \grq^{\btu} \\[1mm]
(\exists x_i \; \grq)^{\btu} 
	& \; \equiv \; & \exists z_i, s_i \;\; \grn[z_i,s_i]
		\;\; \wedge \;\; \grq^{\btu}
\end{array}\end{equation}

%
%

Let $\calN_\times$ be the standard model of \bxbf{PA}, with
zero, successor, addition, and multiplication as functions.

\bthm\label{thm:Arith-to-Structure-sound-complete}
The interpretation $\btu$ is semantically sound and complete:
for every closed formula \grf\ of \bxbf{PA}, $\grf$ is true
in $\calN_\times$ iff $\grf^{\btu}$ is true 
for all $A$-structures.
\ethm
\prf
Consider the following fixed interpretation of the natural numbers
as $A$-structures.
Let $z_0$ be an atom (to interpret 0),
and $s$ an injective unary partial-function 
$z_0 \ara{s} a_1 \ara{s} a_2 \cdot\cdot\cdot$.
That is, $z_0$ and $s$ form a copy of \dN\ in $A$.
Define the {\em canonical} interpretation of $n \in \dN$ 
to be the $A$-structure $(z_0,s_n)$ where $s_n$ is $s$ truncated to 
its first $n$ steps.
Thus every $A$-structure satisfying \grn\ is isomorphic to $(z_0,s_n)$
for some $n$.

We prove that for each formula $\grf[x_1,\ldots , x_k]$ of
\bxbf{PA}, the following conditions are equivalent.
\be
\li $\calN_\times, \; [x_1 \sla n_1 , \ldots , x_k \sla n_k] \models \grf$
\li $\grs \models \grf^{\btu}[z_1,s_1 \ldots z_k,s_k]$ where
\grs\ assigns to $(z_i,s_i)$
the canonical interpretation of $n_i$ ($i=1..k$).
\li $\grs \models \grf^{\btu}[z_1,s_1 \ldots z_k,s_k]$ 
for every \grs\ that assigns to each
$(z_i,s_i)$ an interpretation satisfying \grn\ and of size $n_i$.
\ee

We use induction on \grf.
For the induction basis, (1) and (2) are equivalent by the definition of $\btu$,
(2) implies (3) since the structures
satisfying \grn\ and of size $n$ are all isomorphic to the canonical structure
for $n$, and (3) implies (2) trivially.
The cases for propositional connectives are all immediate.
Finally, for quantifiers we have
that 
$\calN_\times, \; [x_1 \sla n_1 , \ldots , x_k \sla n_k] \models \forall x_{k+1} \grf$
iff 
$\calN_\times, \; [x_1 \sla n_1 , \ldots , x_k \sla n_k, x_{k+1} \sla n] 
		\models \grf$ 
for all $n \in \dN$, which by IH is equivalent to  
$\grf^{\btu}[z_1,s_1 \ldots z_k,s_k,z_{k+1},s_{k+1}]$ being true where $(z_i,s_i)$
is the canonical interpretation of $n_i$ ($i \leq k$)
and $(z_{k+1},s_{k+1})$ is any structure of size $n$ that satisfies \grn,
i.e. $\forall z_{k+1},s_{k+1} \; \grn[z_{k+1},s_{k+1}] \ra \grf^{\btu}$ being
true, i.e.\ $(\forall x_{k+1}\, \grf)^{\btu}$.  This proves the equivalence
of (1) and (3).  The equivalence of (1) and (2) is similar.
\qed

\subsection{Interpreting \bxbf{PA} in \bxbf{FS}}

We proceed to show that the interpretation $\btu$ above from the
language of \bxbf{PA} to that of \bxbf{FS} is not only semantically
sound and complete, but is also a proof-theoretic
interpretation of Peano Arithmetic in the theory \bxbf{FS}.

\bthm\label{thm:PAtoFS}
For every closed formula of \bxbf{PA}, if
\quad $\bxbf{PA} \vdash \grf$, then \quad $\bxbf{FS} \vdash \grf^{\btu}$.
\ethm
\prf
We verify that for every axiom \grq\ of \bxbf{PA},
the formula
$$
\grq^{\Diamond} \;\; \equiv \;\;
	\sbwedge\{\grn[z_i,s_i] \mid \text{$x_i$ free in \grq} \}
	\;\; \ra \;\; \grq^{\btu}
$$
is provable in \bxbf{FS}.

\bi
\li
For Peano's Third Axiom, $\grq \; \equiv \; \tts x_i \neq 0$,  we have
$$
\grq^{\Diamond}\quad  \equiv \quad 
\grn[z_i,s_i]
	\;\; \ra \;\; 
	(\exists t \;\;\bxrm{Suc}[s_i;t] \;\; \ra \;\;
                 \neg \bxrm{Iso}[z_i,t;z_i,\bemptyset])
$$
Reasoning within \bxbf{FS}, 
$\bxrm{Iso}[z_i,t;z_i,\bemptyset])$ implies $t = \bemptyset$,
which contradicts $\bxrm{Suc}[s_i;t]$.

\li
Peano's Fourth Axiom, 
$$
\grq \quad \equiv \quad \tts x_i = \tts x_j \; \ra \; x_i = x_j
$$
is rendered by
\begin{equation*}
\begin{array}{ll}
\grq^\Diamond \quad \equiv \quad & 
	\grn[z_i,s_i] \wedge \grn[z_j,s_j] \\
& \qquad\;\; \ra \;\;
	\bxrm{Suc}[s_i,s_i'] \wedge \bxrm{Suc}[s_j,s_j'] \\
& \qquad\qquad
	\ra \;\; \bxrm{Iso}[z_i,s_i';z_j,s_j']
	\;\; \ra \;\; \bxrm{Iso}[z_i,s_i;z_j,s_j]
\end{array}\end{equation*}
which is easily provable from the definitions.

\li
The defining equations of addition and multiplication are
treated similarly.

\li
Finally, consider an instance of \bxbf{PA}'s Induction Schema,
$$
\grq \quad \equiv \quad
	(\forall x_i  \; \grf[x_i] \sra \grf[\tts x_i])
		\ra \grf[0] \ra \grf[x_j]
$$
with free variables, say a single variable $x_m$.
We have
\begin{equation}\label{eq:ind-delta} 
\begin{array}{ll}
\grq^\Diamond \quad \equiv \quad & \grn[z_m,s_m] \\
& \qquad \ra
	( \;\forall\, z_i,s_i \; \grn[z_i,s_i] \;\wedge\; \grf^\Delta[z_i,s_i]\\
& \qquad\quad
	\wedge \;\; \forall \, t,z_k,s_k \; \bxrm{Suc}[s_i,t]
		\;\wedge \; \bxrm{Iso}[z_i,t;z_k,s_k]
		\;\ra \; \grf^\Delta[z_k,s_k]\; ) \\
& \qquad\qquad \ra \; \grf^\Delta[z_j,\bemptyset] \\
& \qquad\qquad\quad \;\ra \; 
		\grn[z_j,s_j] \;\ra \; \grf^{\btu}[z_j,s_j]
\end{array}\end{equation}
\ei

We shall use the provability in \bxbf{FS}
of the following simple observations:
\be
\li $\grn[z,s] \wedge \bxrm{Iso}[z,s;z',s'] \; \ra \; \grn[z',s']$
\li $\grn[z,s] \wedge \bxrm{Suc}[s;t] \; \ra \; \grn[z,t]$
\li $\grn[z,\{a \mapsto b\} s] \; \ra \; \grn[z,s]$

\li $\grf[z,s] \wedge \bxrm{Iso}[z,s;z',s'] \ra \grf[z',s']$
\li 
$ a,b \not\in f,\vec{g} \; \ra \;
\grx[\{\vec{u} \mapsto a\}\,f,\vec{g}]  \; \ra \;
\grx[\{\vec{u} \mapsto b\}\,f,\vec{g}]$
\ee
To prove $\grq^\Diamond$ within \bxbf{FS} assume
\be
\setcounter{enumi}{5}
\li $\grn[z_m,s_m]$
\li $\grf^\Delta[z_j,\bemptyset]$  
\li $\forall z_i, s_i  \; \grn[z_i,s_i] \wedge \grf^\Delta[z_i,s_i]
                \wedge \forall t,z_k,s_k \; \bxrm{Suc}[s_i,t]
                        \wedge \bxrm{Iso}[z_i,t;z_k,s_k]
                        \;\ra \; \grf^\Delta[z_k,s_k])$
\li $\grn[z_j,s_j]$
\ee
We use f-Induction for the formula 
$$
\grz[s] \quad \equiv \quad \grn[z_k,s] \ra \grf^{\btu}[z_k,s] 
$$
By (7) and (1) we have $\grz[\bemptyset]$.
Assume $\grz[s]$, and consider $\grz[s']$ for $s' = \{a\mapsto b\}\,s$.
If  $\grn[z_k,s']$, then  $\grn[z_k,s]$,
by (3), and so $\grf^{\btu}[z_k,s]$, from $\grz[s]$.
This implies, by (8), 
$$
\forall t,z_k,s_k \; \bxrm{Suc}[s,t]
                        \wedge \bxrm{Iso}[z,t;z_k,s_k]
                        \;\ra \; \grf^\Delta[z_k,s_k])
$$
But we have $\bxrm{Suc}[s;s']$, so taking $s'$ for both $s_k$ and $t$, 
and $z$ for $z_k$, we get $\grf^\Delta[z,s']$.
We have thus completed both the basis and the step of 
the f-induction for $\grz[s]$, and conclude $\grq^\Diamond$.

\medskip

In summary, we have shown that for every formula \grf\ of \bxbf{PA} we have
\begin{equation}\label{eq:Delta}
\bxbf{PA} \vdash \grf \qquad \text{implies} \qquad
	\bxbf{FS} \vdash \grf^\Delta
\end{equation}
\qed

\medskip

We shall show in Theorem \ref{thm:faithful} below that the inverse
of (\ref{eq:Delta}) also hold, i.e. the interpretation $\grf \mapsto \grf^{\btu}$
is {\em faithful.}

\subsection{Interpretation of \bxbf{FS} in \bxbf{PA}}

Developing finite mathematics (and even much of mathematical analysis
\cite[Part 2]{Takeuti78}  
in \bxbf{PA} 
is nowadays a routine exercise.
It would still be useful to articular a particular interpretation. 
To simplify, we refer to the extension $\overline{\bxbf{PA}}$  of \bxbf{PA}
with identifiers for all primitive recursive functions and with their 
defining equations as axioms.
$\overline{\bxbf{PA}}$ is well-known to be conservative over \bxbf{PA}
\cite{Troelstra73}.

Our interpretation $\grq \mapsto \grq^{\btd}$ of \bxbf{FS} into \bxbf{PA}
represents the atoms by the natural numbers:
posit an enumeration $a_0,a_1 \ldots$ of the atoms, and
represent $a_n$ by $n\splus 1$. $\bot$ is represented by $0$.
We code a finite set $S = \{m_1, \ldots, m_k\}$ of natural numbers, displayed
in increasing order,
by the natural number $\# S$ whose binary numeral is $1d_0 \ldots d_{m_k}$,
where $d_i$ is 1 iff  $i \in S$.  Thus $\emptyset$ is coded by 1,
$\{0\}$ by 3 (binary 11), and $\{0,2\}$ by 13 (binary 1101).
We represent an entry $e = \lng a_{n_1}, \ldots , a_{n_k},a_m\rng$ of 
a $k$-ary function ($k>0$)
by the number $\# e$  whose binary numeral is 
$10^m10^{n_1} 1 \cdots 1 0^{n_k}$.
A $k$-ary $A$-function $f$ is the finite set of its entries,
so we represent it as $\# f = \# \{ \# e \mid
	\text{$e$ an entry of $f$} \; \}$.

Clearly we have for each $k \geq 0$ a primitive-recursive (p.r.) predicate
$G_k$ that identifies the codes of strict $k$-ary $A$-functions.
We use these predicates to refer to the syntax of \bxbf{FS},
and to interpret formulas \grq\ of \bxbf{FS} as formulas $\grq^{\btd}$ 
of (extended) \bxbf{PA}, with quantifiers bounded to their intended range.

\bprop\label{prop:flat-semantics}
For every formula \grq\ of \bxbf{FS},
\grq\ is valid iff $\grq^{\btd}$ is true in the structure
$\calN_{{\tiny{\boldmath \!P\!R}}}$\ of the natural numbers with the primitive recursive functions.
\eprop
\prf  By straightforward induction on \grq.
\qed

The semantic correctness of the interpretation $\grq \mapsto \grq^{\btd}$
extends to a proof theoretic interpretation:

\bprop\label{prop:flat-FStoPA}
For every formula \grq\ of \bxbf{FS},
if \quad $\bxbf{FS} \vdash \grq$ \quad then \quad 
$\bxbf{PA} \vdash \grq^{\btd}$.
\eprop
\prf
Other than f-Induction, the interpretations of the axioms of \bxbf{FS} 
are all trivially provable in \bxbf{PA}. As for an instance
$\grq$ of f-Induction (with fixed arity), 
$\grq^{\btd}$ is provable in \bxbf{PA} by induction on the size of the
$A$-function $f$.
\qed


From Proposition \ref{prop:flat-FStoPA} we infer:

\bthm\label{thm:faithful}
The interpretation $\grf \mapsto \grf^{\btu}$ is faithful.  That is,
for every closed formula \grf\ of \bxbf{PA},
\begin{equation}\label{eq:Delta-faithful}
\bxbf{FS} \vdash \grf^\Delta
	\qquad \text{implies} \qquad
		\bxbf{PA} \vdash \grf 
\end{equation}
\ethm
\prf
%

Proposition \ref{prop:flat-FStoPA} implies that for a \bxbf{PA}-formula \grf,
\begin{equation}\label{eq:FS-to-PA}
\bxbf{FS} \vdash \grf^{\btu} \quad \text{implies} \quad
	\bxbf{PA} \vdash (\grf^{\btu})^{\btd}
\end{equation}
It remains to observe that for every \bxbf{PA}-formula \grf,
\begin{equation}
\bxbf{PA} \; \vdash \; (\grf^{\btu})^{\btd} \; \ra \; \grf
\end{equation}
This is proved by induction on \grf, starting with the five forms of
(\ref{eq:5equations}).  The details are straightforward.
\qed

\small

\bibliographystyle{plain}
\bibliography{x}
\end{document}